\documentclass[12pt,preprint]{elsarticle}

\usepackage{amssymb,amsmath}
\usepackage{color}
\usepackage{physics}
\usepackage{centernot}
\usepackage{siunitx}

\providecommand{\vect}[1]{\pmb{#1}}

\usepackage{booktabs}

\usepackage{svg}

\journal{Journal of Magnetism and Magnetic Materials}

\begin{document}

\begin{frontmatter}



\title{Macroscopic emulation of microscopic magnetic particle systems}


\author[LU]{Viesturs Spūlis}
\author[LU]{Daniels Gorovojs}
\author[LU]{Jānis Pudāns}
\author[LU]{Rolands Lopatko}
\author[LU]{Andris P. Stikuts}
\author[LU]{Mārtiņš Brics}
\author[LU]{Guntars Kitenbergs}
\author[LU]{Jānis Cīmurs}
\ead{janis.cimurs@lu.lv}


\affiliation[LU]{organization={MMML lab, Department of Physics, University of Latvia},
            addressline = {Jelgavas~iela~3}, 
            city={Riga},
            postcode={LV-1004}, 
            country={Latvia}}

\begin{abstract}
In this work we show that macroscopic experiments can be used to investigate microscopic systems. Such macroscopic experiments enable testing the assumptions and results obtained by theoretical considerations or simulations that can not be obtained under microscope (e.g., the orientation of a spherical particle).

To emulate the dynamics of a single hematite cube immersed in water and subjected to a rotating magnetic field, a cubic magnet is firmly positioned in a 3D printed superball shell. The dimensionless parameters of the microscopic and macroscopic systems can be equalized by using a glycerol-water mixture instead of water as well as by the material and infill of the 3D printed shell. The pose of the superball is tracked using ArUco stickers which act as fiducial markers.

It was found that there is a qualitative agreement between the macroscopic experiments and theoretical predictions. However, the reduced thermal effects in the macroscopic experiments lead to an increase in friction between the superball and the surface, thus shifting the critical frequency of the system.

Since the shell is 3D printed, the given method can be extended to any shape and magnetization orientation. 
\end{abstract}

\begin{graphicalabstract}
\includegraphics[width=\columnwidth]{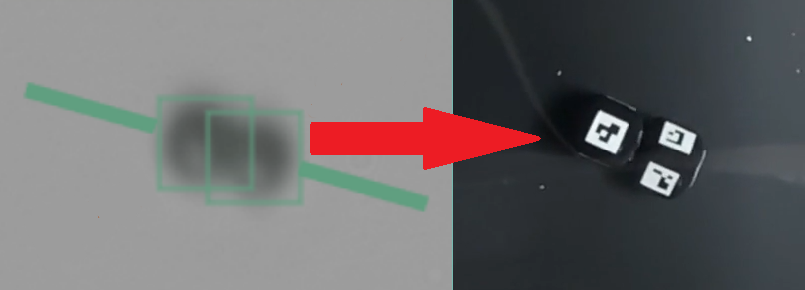}
\end{graphicalabstract}

\begin{highlights}
\item Microscopic magnetic materials can be emulated by small magnets in 3D printed shells
\item Macroscopic emulation reveals complex dynamics in higher detail
\item With macroscopic experimental setup it is possible to observe all motion modes of a microscopic hematite cube. 
\end{highlights}


\end{frontmatter}


\section{Introduction}
\label{sec:intro}

The idea of building analogue macroscopic experiments to validate theoretical results originally derived for microscopic systems has a history dating back at least a century. For instance, in the classical work by G. B. Jeffery, he calculates the increase of viscosity of a fluid containing small elliptical particles \cite{jeffery1922motion}. He proposes a hypothesis about the equilibrium orientation of the particles, which could not be evaluated by the vanishing inertia theory.
Later G. I. Taylor validated the hypothesis by using millimeter sized particles and a macroscopic apparatus filled with a highly viscous fluid \cite{taylor1923motion}.
Similarly, Taylor developed an approximate theoretical description of the viscosity of emulsions of tiny droplets \cite{taylor1932viscosity}.
Afterwards, he validated the theory and found its limits using a macroscopic experiment, where he visually observed the deformation of individual centimetre sized drops in a high-viscosity fluid \cite{taylor1934formation}.

Some recent works using macroscopic experiments focus on the behaviour of many-particle systems -- for example, 3D self-assembly using dipolar interactions in turbulent flow \cite{Abelmann2020}, self-assembly in a suspension on the millimeter scale \cite{miliSelfAssembly}, and self-assembly based on macroscale tiles \cite{MacroSelfAssembly} to name just a few.

In order to better understand the behavior of a microscopic multi-particle system, it is desirable to examine what is happening at the individual particle level. 
However, the orientation of isotropic (spherical) particles \cite{Tierno2023,Tierno2022} and even of colloidal microscopic cubic particles \cite{Brics2022, Brics2023, Cube2021}, as well as particles of other shapes, can only be seen in a limited way under an optical microscope. 
The latter is because the resolution is insufficient to determine the orientation of the particle. 
The resolution can be improved by using an electron microscope \cite{QElectronmicroscope2016,Jin2021,TEM2021} and the particle orientation can be obtained using nitrogen-vacancy centers \cite{Hall2010,NV-centerKaspars}. However, both methods require stationary particles and thus do not allow investigating the dynamic behavior of the particles.

Numerical simulations are widely used to describe the discrepancies between the theory and microscopic experiments.
For instance, they allow researchers to find regimes that were missed in the theoretical evaluation. 
However, not all discrepancies between theory and experiment can be explained using simulations because simulations might contain the same faulty assumptions which do not match the experimental realization.
Numerical simulations can also be used to gain deeper insight into the experiment by adding all relevant physics in the simulation. If the resulting simulation behavior coincides with experiment,  simulation results can give the dynamics of parameters that are not observable in the experiment. These simulations are usually  time consuming due to complicated numerical models.

The advantage of the macroscopic measurements is that it is possible to observe the behavior of the system with the naked eye and from any angle. 
This allows the researcher to determine the movement of all particles and grasp the dynamics at a much higher level of detail. 
However, when comparing a macroscopic system with the analogous microscopic system, it is important to equate the relevant dimensionless quantities.
The cost of enlarging the particles in the macroscopic experiment is an increase in the Reynolds number (for reasonable rotation speeds) and a decrease in thermal noise effects. 
Other drawbacks include limiting the number of particles that can be investigated, a sizable experimental setup and difficulties with generating corresponding external fields (e.g., the magnetic field).
The ability to investigate large ensembles of particles is important, as systems of colloidal microscopic cubic particles have interesting collective effects, including swarming\cite{PETRICHENKO2020} and behaving as a chiral fluid \cite{Soni}.

This work has been motivated by the limitations in investigating the dynamics of a small number of microscopic hematite cubes, reported in Brics et al. \cite{Brics2022,Brics2023}.
There the resolution limitation of the optical microscope did not allow us to study the dynamics of a single hematite cube in a rotating field. 
Here we overcome it by emulating a single hematite cube as a macroscopic superball, as shown below.

In section 2 physical mechanisms of a magnetic superball (hematite cube) in a viscous liquid in a rotating magnetic field are explained and dimensionless parameters relevant to this motion are introduced. In section 3 the physical setup is presented including fabrication of macroscopic magnetic superball. In section 4 the image processing algorithm to obtain superball pose is explained. In section 5 the results of single superball in a rotating magnetic field are shown including phase diagram and orientation tracking. The main findings are summarized in section 6.

\section{Theory}
\label{sec:theory}

\begin{figure}
    \centering
    \includegraphics[width=0.95\columnwidth]{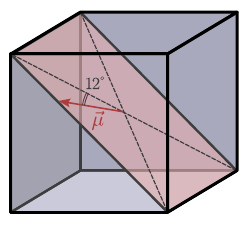}
    \caption{The direction of the magnetic moment \(\vec{\mu}\) in relation to the hematite cube. The magnetic moment lies on a plane defined by two body diagonals. It forms an angle of \(12^\circ\) with one of the body diagonals and points out of the face of the cube. Here the superball-shaped hematite particle is schematically represented as a cube.}
    \label{fig:cube-magnetic-moment}
\end{figure}

The shape of a magnetic hematite cube closely resembles a superball, the surface of which is given by Eq.~\ref{eq:superball}. 
The shape parameter $q$ defines how close the superball is to a perfect cube, with the typical $q=\left[1.5,2\right]$ for hematite cubes, while the constant $a$ is the edge length \cite{Brics2022}. In the following analysis, the value of $q=2$ will be used.
\begin{equation}\label{eq:superball}
    \left|\frac{2x}{a}\right|^{2q}+\left|\frac{2y}{a}\right|^{2q}+\left|\frac{2z}{a}\right|^{2q}=1
\end{equation}
The magnetic moment of a hematite cube is located on the plane defined by two body diagonals, at an angle of $12^\circ$ against one of the diagonals such that \(\vec{\mu}\) is pointing out of a cube face \cite{Brics2022} (see fig. \ref{fig:cube-magnetic-moment}).

The motion of a cube in a rotating magnetic field can be modeled using a microscopic model \cite{Brics2023} of a magnetic superball. The equations of motion (EOMs) are based on the Newtonian mechanics approach where only the essential forces are introduced (hydrodynamics force $\vec{F}^\text{HD}$, buoyancy force $\vec{F}^\text{b}$, and reaction force with bottom of the capillary $\vec{F}^\text{wall}$). This keeps the number of parameters minimal and eases interpretation.

For a hematite cube with an edge length $a$,  magnetic moment $\vec{\mu}$, mass $m$, and moment of inertia tensor $\vect{I}$  in an external homogeneous magnetic field $\vec{B}$ the force and torque balance reads as follows:
\begin{align}
 &{m\frac{\mathrm{d}}{\mathrm{d} t}\vec{v}}=\vec{F}^\text{HD}+
      \vec{F}^\text{b}+\vec{F}^\text{wall},\label{eq:forces}\\
      &{\vect{I}\frac{\mathrm{d}}{\mathrm{d} t}\vec{\Omega}}=\vec{\mu}\times{\vec{B}}+ \vec{T}^\text{HD}+
      \vec{T}^\text{wall},
      \label{eq:torques}
\end{align}
 where $\vec{\mu}\times{\vec{B}}$ is the magnetic torque produced by the external field and $\vec{T}^\text{HD}$,  $\vec{T}^\text{wall}$ are torques of corresponding forces in  Eq.~\ref{eq:forces}. Since the center of mass coincides with the center of the geometric shape, the buoyancy force does not induce torque on the superball.

Unfortunately, the equation system Eqs~\ref{eq:forces}--\ref{eq:torques} is not closed as the knowledge of $\vec{\Omega}$ at a specific time is not sufficient to determine the orientation of the superball. Thus, reaction forces and torques with the bottom surface can not be calculated. To overcome this issue quaternions $\vec{q}$ are used \cite{quat}. Quaternions themselves are four dimensional normalized quantities which provide a convenient representation of spatial orientations and rotations of elements in three dimensional space.  Quaternions, in principle,  correspond to a rotation matrix, however, the quaternion approach requires us to calculate the time evolution of lower dimensional quantity than rotational matrices. Equations of motion for the quaternions
\begin{equation}
 \frac{\mathrm d}{\mathrm{d} t} \vec{q}=\frac{1}{2}\begin{pmatrix}
0 & -\Omega_z&  \Omega_y &  \Omega_x \\
\Omega_z & 0 & -\Omega_x & \Omega_y \\
-\Omega_y  & \Omega_x & 0 & \Omega_z  \\
-\Omega_x & -\Omega_y & -\Omega_z & 0 
\end{pmatrix} \vec{q} =\vect{Q}(\vec{\Omega})\vec{q}
\label{eq:quatx}
\end{equation}
are always stable \cite{Omelyan1998,Kou2018}, unlike EOM for Euler angles. Note that in this work, the angular velocities are given in the laboratory reference frame, whereas for the quaternions, the scalar part of the latter notation is used \cite{eigen}.  In the code Quaternion operations are  performed using a C++ template library for linear algebra Eigen \cite{eigen}.

 The buoyancy force can be calculated as:
 \begin{equation}
  \vec{F}^\text{b}=d(\rho_\text{s}-\rho_\text{f})g a^3,   
 \end{equation}
 where $g=9.81\,\mathrm{m/s^2}$ is the free-fall acceleration, $\rho_\text{s}$ and $\rho_\text{f}$ are the densities of the particle and fluid respectively and $d=\frac{\sqrt{2}\pi\Gamma^2(5/4)}{3\Gamma^2(3/4)}\approx0.81$ is a prefactor for the volume of the superball $V=d\cdot a^3$(with $q=2$). In both the microscopic and macroscopic cases (see Tab.~\ref{tab1}) the superball particles sediment.

Both for the experimental microscopic \cite{Brics2023} and macroscopic conditions, it turns out that the corresponding Reynolds numbers $Re$ is small and inertial terms can be neglected as (${m\frac{\mathrm{d}}{\mathrm{d} t}\vec{v}}\ll\vec{F}^\text{HD}$) and the relaxation  frequency $f_\text{rel}=\frac{\zeta}{I}$ is several orders of magnitude higher than the driving frequency of the system ($f_\text{rel}\gg f$ see Tab.~\ref{tab1}). 
Therefore,  the Stokes approximation of assuming inertialess particles is used. 
For hydrodynamics forces and torques, to keep the equations analytically tractable, the linear velocity drag approximation is used, which for a cubic shaped particle reads \cite{Okada2018}: 
\begin{align}
 {\vec{F}^\text{HD}}&=-\xi \vec{v}; \quad \xi\approx3 \pi \eta a \cdot 1.384;\\
 {\vec{T}^\text{HD}}&=-\zeta \vec{\Omega} ;\quad  \zeta= \pi \eta a^3 \cdot 2.552,
\end{align}
where $\eta$ is the viscosity of the solvent, and $\xi$ and $\zeta$ are the drag and rotational drag coefficients, respectively. 

The aforementioned reaction forces $\vec{F}^\text{wall}$ and corresponding torques $\vec{T}^\text{wall}$ are added so that the cube does not fall through the bottom surface. The results do not depend on the choice of the exact expression for the reaction forces whenever the model for the reaction forces is reasonably chosen. 
    
Combining all expressions the EOM for dimensionless variables reads:
\begin{align}
\begin{split}
 \tilde v_z&=kS\left(\tilde{F}^\text{b}_{z}+{\tilde{F}}_{z}^\text{wall}\right),\\
   \vect{\tilde \Omega}&=\vect{\hat{\mu}}\times \vect{\hat B}+S\vect{\tilde{T}}^\text{wall},\\
   \frac{\mathrm d}{\mathrm{d} \tilde t} \vect{q}&=\vect{Q}(\vect{\tilde \Omega})\vect{q}, \label{eq:quat}
\end{split}
\end{align}
with $\tilde{}$ denoting dimensionless variables apart from unit vectors, which are denoted with  $\hat{}$,  e.g. $\vect{r}=r\hat{\vect{r}}$. Note that here nondimensionalization for time $\tilde{t}=\frac{\zeta t}{\mu B}$ and distance $\tilde{\vect{r}}=\frac{ \vect{r}}{a}$  is used leading to dimensionless variables $\tilde{\vect{v}}=\frac{\vect{v}\zeta}{ \mu B a}$ , $\tilde{\vect{\Omega}}=\frac{\vect{\Omega} \zeta }{\mu B}$, $\tilde{\vect{F}}=\frac{\vect{F}}{d(\rho_\text{s}-\rho_\text{f})g a^3}$, and $\tilde{\vect{T}}=\frac{\vect{T}}{d(\rho_\text{s}-\rho_\text{f})g a^4}$.  The EOM has two parameters $S$ and $k$, from which only 
\begin{equation}
 S=\frac{d(\rho_\text{s}-\rho_\text{f})g a}{MB}
\end{equation}
is adjustable in experiments by changing the magnitude of the external magnetic field, where $M=\frac{\mu}{V}$ is mean magnetization of the superball. The parameter
\begin{equation}
  k=\frac{ \zeta}{a^2 \xi}\approx0.614
\end{equation}
represents the drag coefficient ratio, which 
is exclusively dependent upon the particle's shape and not its size.

In the system with multiple magnetic superballs the dimensionless gravitational parameter $G_\text{m}=\frac{4 \pi(\rho_\text{s}-\rho_\text{f})g a^7}{3\mu_0 \mu^2}$ should be added to characterize the system \cite{Brics2023}. The gravitational parameter is the ratio of gravitational force to magnetic force.
The importance of gravity in macroscopic system can be emphasised by sedimentation speed $v_\text{sed}=\frac{d(\rho_\text{s}-\rho_\text{f})g a^2}{3 \pi \eta \cdot 1.384}$ (the values shown in the table).

\section{Experimental setup}
\label{sec:setup}
\begin{figure}[ht]
    \centering
    \includegraphics[width=0.95\columnwidth]{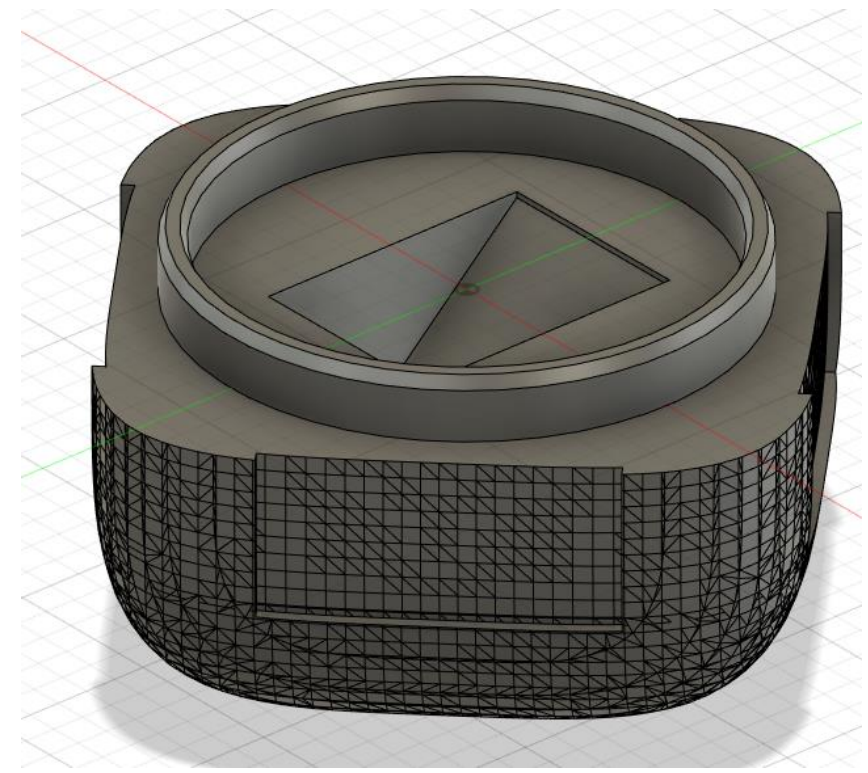}
    \caption{3D model of the superball shell lower half, modelled in Fusion 360 CAD software.}
    \label{fig:superball_Fusion360}
\end{figure}

The macroscopic 3D model geometry was defined with the shape parameters $q=2$ and $a=1\,\mathrm{cm}$. 
A $5\times5\,\mathrm{mm}$ large and $0.1\,\mathrm{mm}$ deep square indent was made on all faces of the superball to allow for accurate placement of markers. 
A cube with an edge length of $3\,\mathrm{mm}$ is subtracted from the center of the superball, allowing for the insertion of a magnet once the shell is printed.
The insert is positioned in such a way as to ensure that the magnetic moment of the cube magnet is positioned in the same way as the magnetic moment of the hematite cube.
The superball is split in half, and circular indents/extrusions are made to connect the two halves of the shell. 
The lower half of the superball model is shown in Fig.~\ref{fig:superball_Fusion360}.
The shells are printed in a resin 3D printer (Elegoo Mars 2 Pro), using black resin for high contrast against the background and markers.
When the lower and the upper parts of the superball have been printed, in the center of the superball we place a 3$\times$3$\times$3 mm neodymium magnet (class N45, \textit{Supermagnete}) with volume magnetization $M=\SI{860}{\kilo\ampere/\meter} $ and magnetic dipole moment $\mu=\SI{0.02}{\ampere \meter^2}$.  
Then the two halves are pressed tightly together. 
On each face of the superball unique ArUco markers \cite{ArUco} are glued. 

\begin{figure}
    \centering
    \includegraphics[width=\columnwidth]{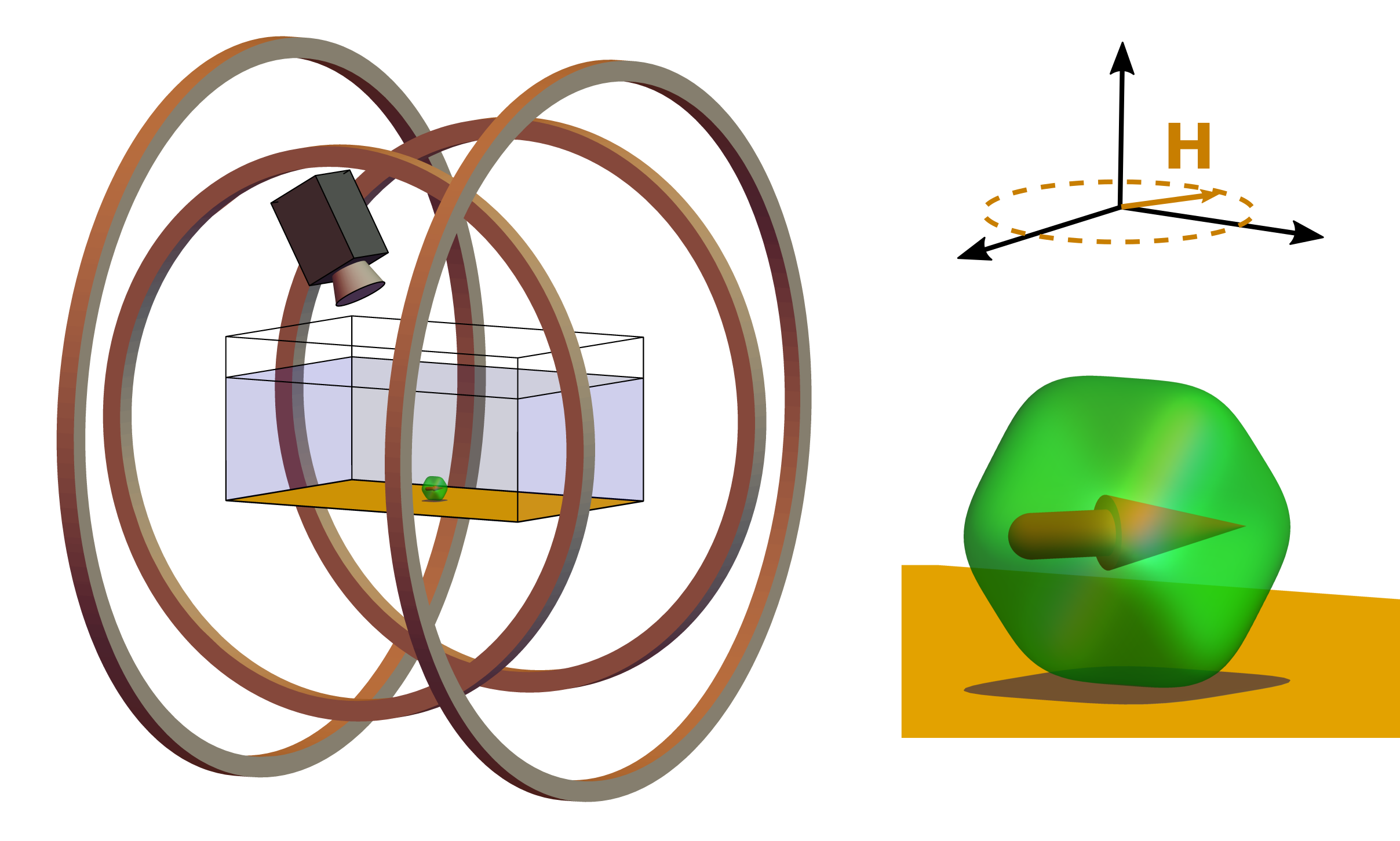}
    \caption{The scheme of the experimental setup: The box filled with glycerol placed within three pairs of Helmholtz coils. Magnetic superball is placed in the box and Helmholtz coils generates in horizontal plane rotating magnetic field.}
    \label{fig:exp-sheema}
\end{figure}

The superball is placed in a rectangular box with inner size L$\times$W$\times$H$=\SI{17}{\centi\meter}\times\SI{11}{\centi\meter}\times\SI{8}{\centi\meter}$. 
The box is half filled with glycerol ($99.5\%$). 
However, the glycerol was left for many days in an open box. 
As a result, due to water absorption, the viscosity of the glycerol-water mixture has decreased to $\eta=\SI{130}{\milli\pascal\second}$, which was measured by Anton Paar Rheometer MCR 502.
The box is placed within three pairs of Helmholtz coils (see \cite{coils} for more details), which generate the magnetic field.
They are powered by 3 current sources (KEPCO), which are controlled by a NI DAQ card using Python code. 
To record the observations we use digital cameras - a mobile phone (Apple) for qualitative measurements and an industrial camera (Basler ac1920-155um) with a camera objective (Nikon) for quantitative measurements using tracking explained in the next section. The scheme of the experimental setup is shown in Figure \ref{fig:exp-sheema}. 

The comparison between parameters of  microscopic experiment (article \cite{Brics2023}) and macroscopic experiment (this paper) can be seen in table~\ref{tab1}.

\begin{table*}[ht]
    \centering
    \begin{tabular}{lccc}
        \toprule
        Parameter& Notation & Micro & Macro \\
        \midrule
        Superball side length & $a$  & \SI{1.5}{\micro\meter}  & \SI{1}{\centi\meter}\\
        
        Viscosity of liquid & $\eta$ & \SI{1}{\milli\pascal.\second} & \SI{130}{\milli\pascal.\second}\\
        
        Mean density of the superball & $\rho_\text{s}$ & \SI{5.25}{\gram/\centi\meter^3} & \SI{1.47}{\gram/\centi\meter^3}\\
        
        Fluid density & $\rho_\text{f}$ & \SI{1}{\gram/\centi\meter^3} & \SI{1.26}{\gram/\centi\meter^3}\\
        
        Mean magnetization of superball & $M$ & \SI{2.2}{\kilo\ampere/\meter} & \SI{31}{\kilo\ampere/\meter}\\
        
        External field strength & $B$ & $\approx\SI{1}{\milli\tesla}$ & $\approx\SI{0.2}{\milli\tesla}$\\
        
        Torque ratio & $S$ & $0.03$ & $0.65$\\
        
        Gravitational parameter & $G_\text{m}$ & $0.04$ & $0.12$\\

        Sedimentation speed & $v_\text{sed}$ & $\SI{5.8}{\micro\meter/\second}$ & $\SI{9.7}{\centi\meter/\second}$\\
        
        Critical frequency of superball & $f_\mathrm{c}$ & $\approx\SI{40}{\hertz}$ & $\approx\SI{0.7}{\hertz}$\\

        Relaxation frequency & $f_\mathrm{rel}$ & $\approx\SI{5}{\mega\hertz}$ & $\approx\SI{60}{\hertz}$\\

        Magnetic field frequency & $f$ &0.1 -- 40$\,\mathrm{Hz}$ & 0.1 -- 10$\,\mathrm{Hz}$\\
        
        Reynolds number & $\mathrm{Re}$ & $\approx 0.001$ & $\approx 4$\\
        \bottomrule
    \end{tabular}
     \caption{Comparison of parameters in microscopic experiments of hematite cubes done in article \cite{Brics2023} and macroscopic experiments of 3D printed superball with enclosed cube magnet presented in this article.}
    \label{tab1}
\end{table*}

\section{Automatic tracking method}
\label{sec:tracking}

The addition of ArUco markers \cite{DetectArUco} on the faces of the superball allows us to track its position and rotation using a camera.
The details of this method are described in this section.

The method of operation of a pinhole camera can be described by a \(3\times4\) matrix which maps 3D coordinates of real-world objects to their position on a 2D image. 
This is often termed the \emph{camera matrix}. 
The camera matrix for a given camera can be determined empirically using multi-plane calibration via chessboard detection \cite{chessboard}. 
This functionality is implemented in various computer visual libraries, such as OpenCV \cite{OpenCV}, which also offers functionality to remove lens distortions from the images.

It is possible to detect and recognise different ArUco markers in an image of the superball \cite{DetectArUco}. 
Each face of the superball has a unique marker, thus it is possible to determine which sides of the superball are facing the camera by considering which markers are detected on the image.

\begin{figure}[ht]
    \centering
    \includegraphics[width=0.95\columnwidth]{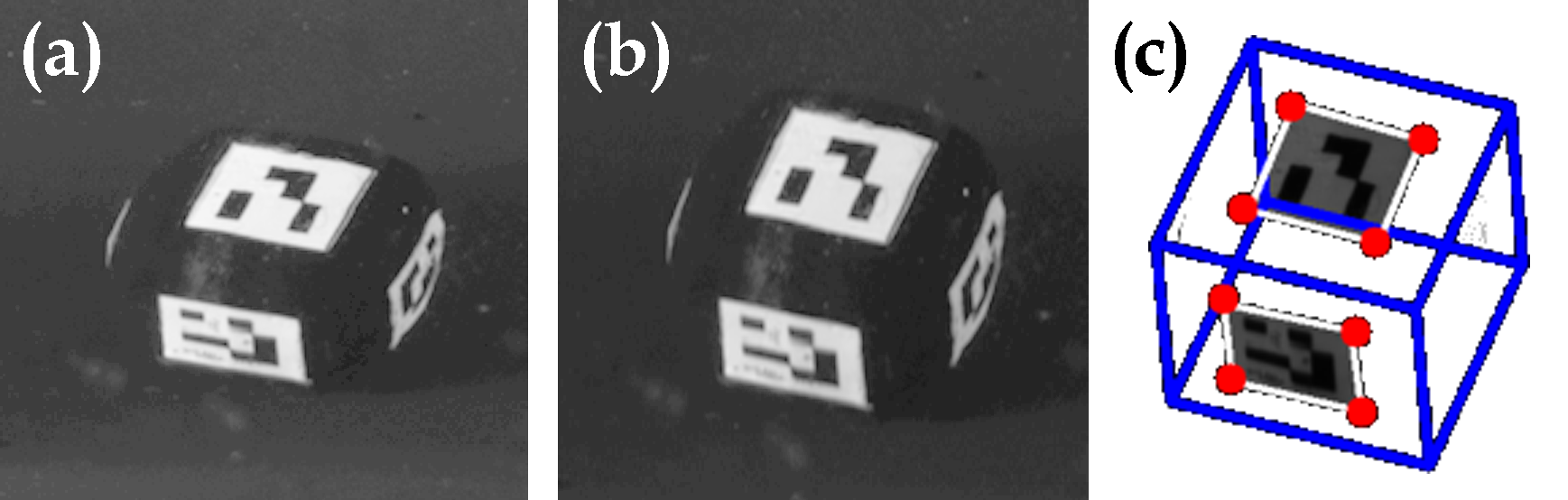}
    \caption{From experimental superball image to detection and pose estimation. (a) Experimental image of a superball. (b) Image after correction for refraction. (c) Estimated pose of the superball (blue contour), where white lines show the detected markers and red dots indicate marker corners. The perspective-n-point algorithm attempts to minimize the distance between the estimated location of the marker corners (red) and the detected markers (white).}
    \label{fig:detected_cube}
\end{figure}

The relative positions between each marker and between their corners on the superball are known. 
By finding the image coordinates of the corners of each marker, it is possible to construct a 2D to 3D point correspondence. 
Then, a perspective-n-point algorithm can be used to recover the position and rotation of the superball by minimising the re-projection error \cite{PnPproblem}. 
An example image with detected markers and estimated superball pose can be seen in Fig.~\ref{fig:detected_cube}.

This method can be extended to track multiple superballs at the same time by ensuring that each superball has a unique set of six markers. 
Having unique markers for each superball allows us to differentiate between each superball in an image. 
However, having too many superballs in the setup is likely to limit the precision to which the position and rotation of superballs can be determined, since some markers might be covered by other superballs.

There are many effects that introduce errors in detecting the coordinates of the corners of the markers. 
These include imperfectly cut and attached markers, optical distortions due to ripples in the glycerol-water mixture, vibrations of the setup and camera, etc. 
There are practical limits to the extent to which these can be addressed, always leaving some errors in corner detection.

To minimize the error in pose estimation, it helps if multiple markers can be detected at once, thus giving more points to use in pose estimation. 
Due to the cube-like shape of the superball, most of the time it will lie flat with one marker facing upwards. Hence, it is inefficient to position the camera right above the setup. 
Instead, if the camera is positioned at an angle to the setup, most of the time at least two markers of a single superball will be visible to the camera. 
This achieves the goal of having more points to use in the perspective-n-point algorithm. 
It should be noted that having the camera at an angle to the liquid-air interface introduces a visual distortion in the images taken due to refraction. 
However, to the first order approximation, this is simply a scaling in two orthogonal directions which can easily be reversed digitally.
For the particular camera position, the example in Fig.~\ref{fig:detected_cube}~(b) includes enlarging (a) image by $30\%$ along the vertical direction. 

Perspective-n-point pose estimation algorithms rely on the perspective distortion caused by some points being closer to the camera than others. 
If the camera is located far from the setup, the perspective distortion is very small. 
Thus, even small errors in the detected positions of marker corners can lead to erroneous rotation and position estimates of the superball. 
This can be minimized by moving the camera closer to the setup and thus causing a greater perspective distortion. 
However, this significantly reduces the field of view.

ArUco markers on a superball allow tracking of its position and rotation as outlined in Sec~\ref{sec:tracking}. Imperfections in determining the exact positions of the markers in an image cause errors in the measured rotation. Aside from small scatter (noise on the order of few degrees), it was observed that occasionally the measured pose was mirrored across a given plane. Often this could be fixed by additional pre-processing of the collected data by guessing the plane across which the pose was mirrored. However, this was not always feasible, as the distortion was too great across the whole measurement. While increased precision in the setup and improved pre-processing methods can likely eliminate this issue, enough data could still be collected.

\section{Results and Discussion}
\label{sec:results}

To validate and justify our macroscopic experimental setup, initially experiments with a superball dimer in a rotating magnetic fields are performed. 
It is shown that, for superball dimers, both synchronous and asynchronous motion of dimer, including breaking and reconnecting, is possible.
This can be seen in the Supplementary Video S1.
In the case of synchronous rotation for magnetic fields, which corresponds to microscopic experiments \cite{Brics2023}, the dimer rotates on an edge with magnetic moment in the plane of the rotating magnetic field, as predicted. 
In the asynchronous regime, back-and-forth rotation is observed. 
However, no modes were observed where the dimer goes out of the plane of the rotating magnetic field. This can be explained by a higher gravitational parameter $G_\text{m}$ in the macroscopic experiments.

Having done this validation, we performed experiments on a single superball, which were not possible in microscopic scale due to microscope resolution limitations \cite{Brics2022, Brics2023}. 
The experiments were divided in two groups. 
First, we tried to visually determine qualitative cube motion characteristics, followed by more precise measurements where motion of each superball's face was tracked. 

\subsection{Phases of a single cube}

The first thing tested in the macroscopic experiment was whether we can visually identify all of the motion modes that were predicted for a single cube in the microscopic experiment \cite{Brics2023}. 
Indeed, in the synchronous regime, when a magnetic particle rotates with the frequency of the rotating magnetic field, we were able to observe all predicted cases in Video2 of \cite{Brics2023}: superball rotation on an edge, corner or face, depending on the rotating magnetic field strength. 
However, rotation on an edge was observable only for small frequencies, which were smaller than predicted in \cite{Brics2023}. 
The reason for that is probably the dry friction between a superball and the bottom of the container which is more pronounced in the macroscopic experiment, where the thermal fluctuations are negligible. 
Thus, in the macroscopic experiment, the superball rotates more frequently on a corner than on an edge and face as friction torque for rotation on the corner is several orders of magnitude smaller.

In the asynchronous regime we were able to observe precession (see Supplementary Video S2) and back-and-forth motion (see lower two videos in Supplementary Video S3). 
However, no complicated motion modes which are combinations of precession and back-and-forth motion as in \cite{Brics2023} were observed. 
This can be explained again with dry friction as precession was observed only for small frequencies slightly above critical frequency when asynchronous motion was observed.

To be more sure that dry friction is the cause for discrepancies, we performed experiments at weak rotating magnetic fields and noted the cube behavior in a phase diagram, which is shown in Fig.~\ref{fig:single_cube_phase_diagram} (see Supplementary Video S3 for some examples).
According to \cite{Brics2023} the superball should always rotate on the face. 
This prediction is true in the synchronous regime, however only for small frequencies. 
In addition, the boundary between rotation on a face and rotation on a corner is almost independent of the magnetic field strength. 
Moreover, the slopes $k_\text{f}$ and $k_\text{c}$ for the lines separating synchronous regimes (rotation on face or corner, respectively) from the asynchronous regime (back-and-forth rotation) are different. 

\begin{figure}[ht]
    \centering
    \includegraphics[width=\columnwidth]{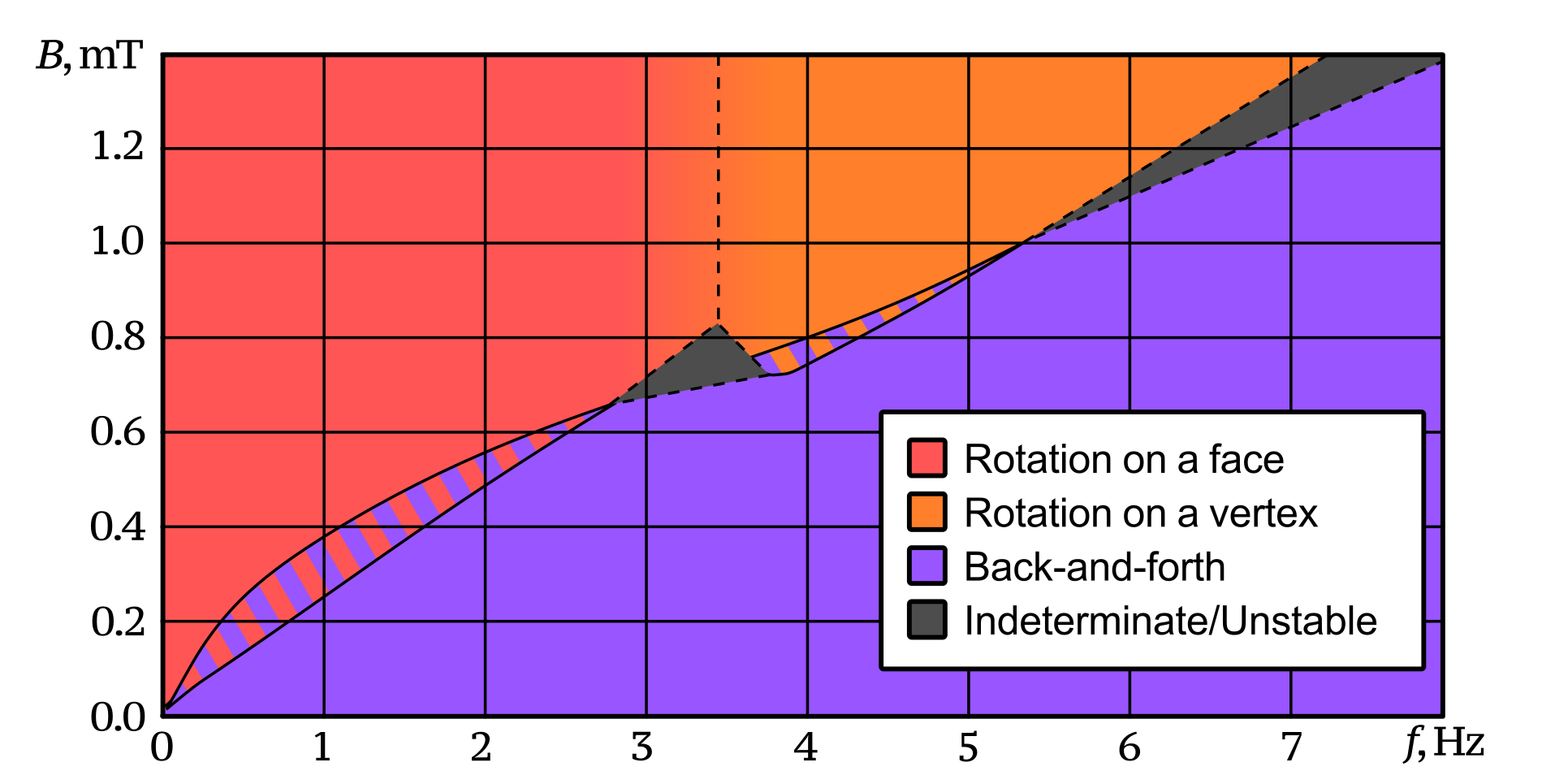}
    \caption{Rough sketch of the observed phases of a single cube from 91 measurements. Bi-colored regions denote magnetic field strength and rotation frequency parameters at which multiple phases were observed to be stable. Dashed lines indicate phase transitions where it was difficult to discern the exact moment when the phase transition occurs.}
    \label{fig:single_cube_phase_diagram}
\end{figure}

If the dry friction is added to the model, the torque balance for synchronous rotation reads:
\begin{equation}
    \zeta \vec\Omega=\vec{\mu} \times \vec{B} -\vec{T}_\mathrm{Fr},
\end{equation}
where $\vec{T}_\mathrm{Fr}$ is torque due to dry friction which is independent on velocity. Thus, for the line separating synchronous regime from asynchronous regime average slope $k_i$ and intercept $b_i$ reads:
\begin{equation}
    k_i=\frac{2 \pi \zeta}{\mu_\mathrm{plane}};\quad b_i =\frac{T_\mathrm{Fr}}{\mu_\mathrm{plane}},
\end{equation}
where $\mu_\mathrm{plane}$  is the projection of the magnetic moment in the plane of the rotating magnetic field. When a superball rotates on a face $\mu_\mathrm{plane}\approx 0.877\mu$, but when on a corner $\mu_\mathrm{plane} \in (0.877\mu, \mu)$. Thus, the line separating the asynchronous regime from the synchronous regime, where the superball rotates on a corner, has $\mu_\mathrm{plane}\approx \mu $.

From Fig.~\ref{fig:single_cube_phase_diagram} for $f>0.5\,\mathrm{Hz}$ for rotation on a face one finds $k_\text{f}\approx (0.2\pm 0.1) \,\mathrm{mT/Hz}$, $b_\text{f}\approx (0.2\pm 0.4)\,\mathrm{mT}$ and for rotation on a corner $k_\text{c}\approx (0.2\pm 0.1) \,\mathrm{mT/Hz}$, $b_\text{c}\approx 0\,\mathrm{mT}$. The ratio $k_\text{c}/k_\text{f}$ matches the predicted value of 0.877 within the margin of error. Therefore, for rotation on a face if 
$f>\SI{0.5}{\hertz}$ the dry friction torque is $T_\mathrm{Fr}\approx (4\pm 1) \,\si{\micro\newton\meter}$ which coincides with rough estimate $T_\text{rough estimate}=\mu (\rho_\text{s}-\rho_\text{f}) V g \frac{a}{2}=\SI{3.8}{\micro\newton\meter}$ for $\mu=0.5$ and $V=d a^3$.

The alternative hypothesis accounting for deviations between experimental observations and theoretical predictions may stem from uncompensated Earth magnetic field. In instances of low magnetic field strength, specifically when the cube rests on its face, only the magnetic component parallel to the ground exerts influence on the results. However, at higher field strengths ($B > \SI{0.7}{\milli\tesla}$), when the cube is positioned on its vertex, the vertical component of the Earth's magnetic field also becomes a contributing factor.

Numerical simulations were conducted to assess the impact of Earth's magnetic field on the critical frequency $f_\text{c}$ at the experimental site in Riga, Latvia, where the Earth's magnetic field strength $B_\text{Earth}$ is $\SI{52}{\micro\tesla}$ with an inclination angle $\theta_\text{Earth}=71^\circ$. The findings revealed that the influence on $f_\text{c}$ remains below 0.2\% across the entire measurement range. This suggests that, in this particular location, the Earth's magnetic field has a negligible effect on the experimental outcomes. It is noteworthy, however, that regions characterized by a smaller inclination angle of the Earth's magnetic field $\theta_\text{Earth}$ may experience a more substantial impact.

\subsection{Tracking markers}

In general, the rotation of a three-dimensional object can be expressed via three Euler angles. However, only one angle is sufficient when the object rotates around a single fixed axis. Thus, the simplest example to track and attempt to fit to theory is the motion of a superball when it rotates around the axis normal to the surface it rests on. This subsection provides such an example of a superball exhibiting back-and-forth motion around a fixed axis.

\begin{figure}[ht]
    \centering
    \includegraphics[width=\columnwidth]{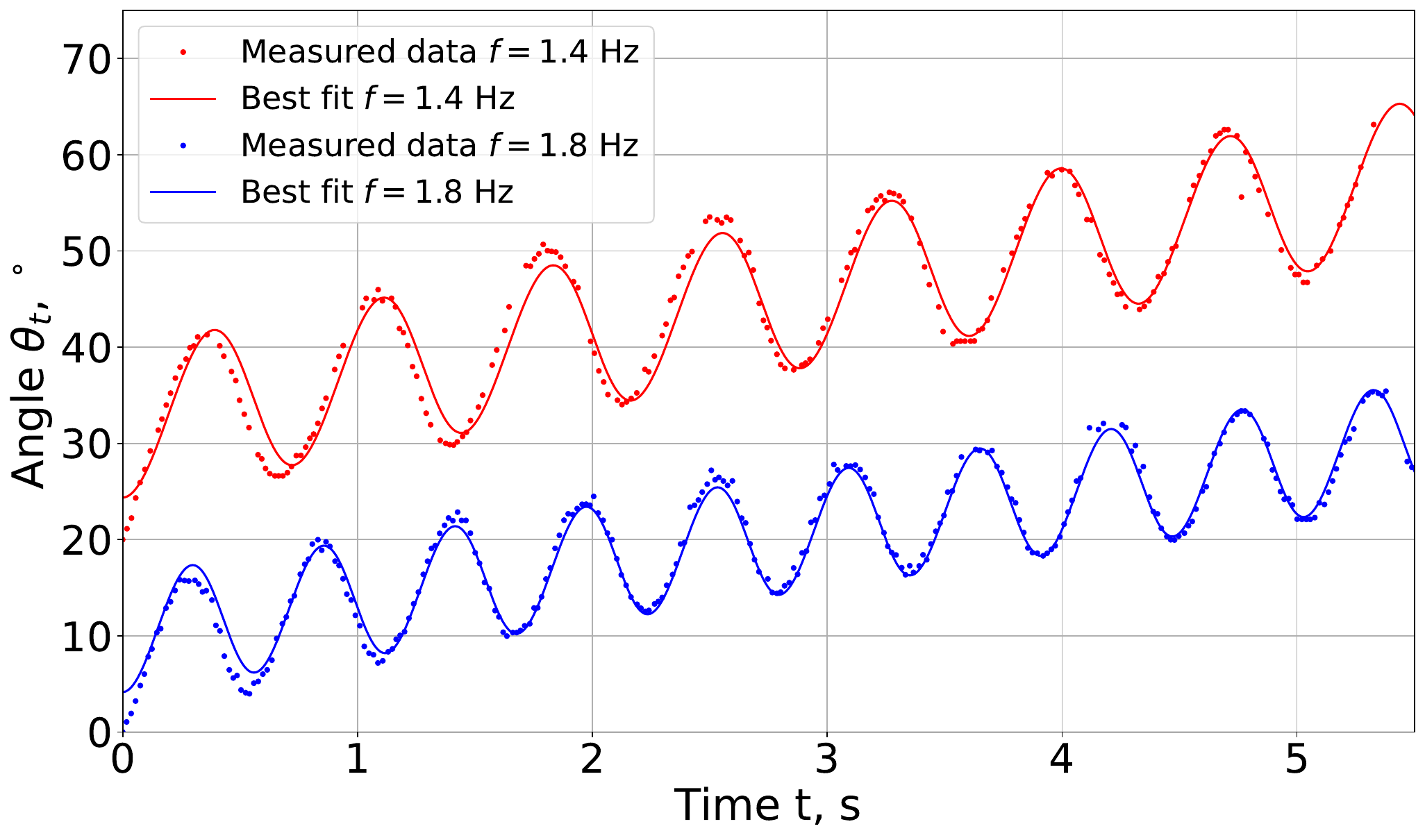}
    \caption{Measured angle of a superball around an axis perpendicular to the surface the superball rests on. The superball is subjected to a magnetic field strength of \(0.15\,\mathrm{mT}\) rotating with frequencies of \(f=1.4\,\mathrm{Hz}\) ($\omega=8.94 \, \mathrm{rad/s}$) and \(f=1.8\,\mathrm{Hz}\) ($\omega=11.30 \, \mathrm{rad/s}$). Solid curves are the fit of Eq.~\ref{eq:curve_fit_function} with the best-fit parameter \(\omega_\text{c}=(1.19\pm0.01)\,\mathrm{rad/s}\) for both frequencies.}
    \label{fig:curve_fitting}
\end{figure}

Fig.~\ref{fig:curve_fitting} shows the measured angle, \(\theta_t\), of a superball in the plane of rotation subjected to a uniformly rotating magnetic field. The field rotates in-plane with the surface on which the superball rests. This data is fit to Eq.~\ref{eq:curve_fit_function}, which is adapted from \cite{Brics2023}. 
The fit, in principle, allows us to extract two crucial parameters---frequency of the rotating magnetic field, \(\omega\) (which can be compared with the coil settings), and the critical frequency when back-and-forth motion starts, \(\omega_\text{c}\) (which can be compared with theory). However, here to be sure that we are not overfitting, since for small times there are some discrepancies between fit and measured data, the $\omega$ was fixed to the value extracted from the coil data. The remaining two variables, \(t_0\) and \(\alpha\), set the initial time and angle offset respectively.

\noindent
\begin{align}
    \theta_t=\omega&\cdot(t-t_0)+\alpha\nonumber\\
    &- 2\arctan\qty[\frac{\omega_\text{c}+\sqrt{\omega^2-\omega_\text{c}^2}\tan\qty(\frac{1}{2}\qty(t-t_0)\sqrt{\omega^2-\omega_\text{c}^2})}{\omega}]\nonumber\\
    &\qquad{}-2\pi\left\lfloor\frac{(t-t_0)\sqrt{\omega^2-\omega_\text{c}^2}}{2\pi}+\frac{1}{2}\right\rfloor
    \label{eq:curve_fit_function}
\end{align}
The first term of Eq.~\ref{eq:curve_fit_function} is the continuous rotation due to the rotating magnetic field, and the second term is an initial angle offset. The third term describes how much the superball ``lags behind'' the magnetic field due to its inability to rotate sufficiently fast. The final term in Eq.~\ref{eq:curve_fit_function} makes the arctangent term continuous, where $\lfloor x\rfloor$ brackets denotes integer part of $x$.

Upon visual inspection of Fig.~\ref{fig:curve_fitting}, it seems that the measured data match the best-fit line rather well, although for small times there are notable discrepancies which most likely are caused by the inertial effects of the fluid. 
The fluid flow was created by the superball, when the external magnetic field was turned on as for rotation with relaxation frequency $f_\text{rel}$ (see Tab.~\ref{tab1}) $Re\gg 1$. 
For later times, the curve fits the measured data notably better. The data were also fitted with a numerical solution of the Eq.~\ref{eq:torques}, including the constant magnitude friction torque calculated in the previous section. The curves were hardly distinguishable from the ones obtained by  Eq.~\ref{eq:curve_fit_function}, therefore they are not shown in Fig.~\ref{fig:curve_fitting}. Thus, it can be concluded that dry friction does not influence the solid curve shape in  Fig.~\ref{fig:curve_fitting}. Only the critical value of the magnetic field is affected.

Although there is some noise and scatter in the measured data, the characteristic back-and-forth motion is clearly discernible and quantitatively consistent with theory. The determined critical frequency $\omega_\text{c}=(1.19\pm0.01)\,\mathrm{rad/s}$ ($f_\text{c}=(0.189\pm0.002)\,\mathrm{Hz}$) within the error range is highly consistent with the critical value determined for $B=0.15\,\mathrm{mT}$ in Fig.~\ref{fig:single_cube_phase_diagram}. This demonstrates that the method of tracking a superball using ArUco markers is viable. While analysis similar to the example in this subsection can be done on a superball undergoing a more complicated motion (e.g., with rotation along multiple axes), such examples are out of scope for this paper.

\section{Conclusions}
\label{sec:coclusions}

Using macroscopic experiments, it is possible to check the theoretical predictions for a single magnetic superball in a rotating magnetic field. Macroscopic experiments allowed us to distinguish between rotation on the face, rotation on the edge and rotation on the corner of a magnetic superball. This allowed us to construct a phase diagram of the single magnetic superball in the rotating magnetic field (Fig. \ref{fig:single_cube_phase_diagram}). Qualitatively, the phase diagram coincides with theoretical predictions. The pose of the superball can be obtained using ArUco markers on each face.

Deviations from theoretical predictions were observed in the macroscopic experiments, which can be explained by dry friction between the superball and the bottom of the container. Further improvements for the given experimental setup are needed to remove the effect of dry friction to better emulate microscopic system where dry friction plays negligible role.

Macroscopic experiments with a 3D printed shell and magnetic core can be used to simulate a wide variety of magnetic micro-particles. Adding water to glycerol enables control over the viscosity and density of the liquid. The size of the magnet, form, material and infill of the 3D printed shell allows one to control the parameters of the particle. This allows equalization of dimensionless parameters with the microscopic system and obtaining critical values in the desired range. For single magnetic superball experiments, only one dimensionless parameter should be equalized.
If the size of the coil system is sufficient, the given approach can be used to investigate many particle systems in higher resolution. For a many magnetic superball system there would be two dimensionless parameters to equalize.

\section{Acknowledgments}
This research is funded by the Latvian Council of Science, PostDoc Latvia project No.1.1.1.2/VIAA/3/19/562 and project No.1.1.1.2/VIAA/1/16/060, and FLPPLV project BIMs No. lzp-2020/1-0149.


 \bibliographystyle{elsarticle-num} 
 \bibliography{references}





\end{document}